\documentstyle[11pt,moriond,epsfig]{article}

\def\cpc#1#2#3{Computer Phys.\ Comm.\ #1 (19#3) #2}

\def\np#1#2#3{Nucl.\ Phys.\ B#1 (19#3) #2}
\def\pl#1#2#3{Phys.\ Lett.\ #1B (19#3) #2}

\def\zp#1#2#3{Zeit.\ Phys.\ C#1 (19#3) #2}

\def\be{\begin{equation}}
\def\ee{\end{equation}}
\def\ba{\begin{eqnarray}}
\def\ea{\end{eqnarray}}
\def\bann{\begin{eqnarray*}}
\def\eann{\end{eqnarray*}}
\def\benn{\begin{displaymath}}
\def\eenn{\end{displaymath}}
\def\nn{\nonumber}

\def\n{\, {\cal N}}
\def\tN{{\cal \widetilde{N}}}

\def\qqb{Q \bar Q}

\newcommand\as{\alpha_{\mathrm{S}}}

\newcommand\C{{\mathrm{c}}}
\newcommand\B{{\mathrm{b}}}
\newcommand\Z{{\mathrm{Z}}}
\newcommand\smfrac[2]{{\textstyle\frac{#1}{#2}}}

\begin{document}
\begin{flushright}
  RAL-TR-98-043 \\ hep-ph/9806302
\end{flushright}
\vspace*{3.4cm}
\title{THE MULTIPLICITY OF HEAVY QUARK PAIRS FROM GLUON SPLITTING IN 
\boldmath$\mathrm{e^+e^-}$ ANNIHILATION}

\author{ D.J. MILLER }

\address{Rutherford Appleton Laboratory, Chilton, Didcot, Oxfordshire.  
OX11 0QX.  England.}

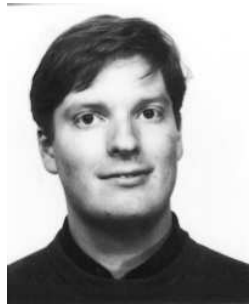
\maketitle\abstracts{
\pretolerance 800
I will briefly review the progress which has been made in the
investigation of secondary heavy quark production at LEP.  I will show
why a calculation of the secondary heavy quark multiplicity, keeping
the dependence on an event shape variable, is essential for better
exploitation of data, and I will present the results of such a
calculation. This will be compared to Monte Carlo studies.}

\section{Introduction}

Heavy quark production in $\mathrm{e^+e^-}$ annihilation can come from
two possible sources: from the hard interaction itself,
$\mathrm{e^+e^- \to \qqb}$, and from the splitting of perturbatively
produced gluons, $\mathrm{e^+e^- \to q \bar q g \to q \bar q \qqb}$. I
shall refer to the former as {\em primary} heavy quarks and the latter
as {\it secondary} heavy quarks.

Of course, in order to be able to regard primary and secondary heavy
quark production as separate processes, the interference between them
must be zero, or at least very small. Fortunately this is the
case[\ref{Mike}]. For non-identical quarks coupled to a vector
current, the interference between them will vanish by Furry's theorem
if we assume that the charges of the quarks are not measured. For an
axial current, cancellations will occur between up and down type
quarks, leaving only the case where the ``light'' quark is a bottom
quark. This will only provide an effect of the order of $0.2\%$ of the
secondary heavy quark rate[\ref{Kniehl}]. The contribution for
identical quarks is only slightly larger. In this case Furry's theorem
no longer applies but necessarily the quarks will be of the same
flavour (bottom or charm) and subsequently give only a small
contribution. Therefore one can regard secondary heavy quark
production as a separate process and investigate it both
experimentally and theoretically.

In this talk, I will briefly review the progress which has been made
in this investigation. I will
show why a calculation of the secondary heavy quark multiplicity,
keeping the dependence on an event shape variable, is essential for
better exploitation of data, and I will present the results of such a
calculation. This will be compared to Monte Carlo studies.

\subsection{Theoretical Progress}

The heavy quark mass provides a natural infrared cut-off for the
theoretical calculation of the total rate of secondary heavy quark
production. It is therefore an infrared-safe quantity, and can be
calculated as an order-by-order perturbative expansion in $\as$,
starting at ${\cal O}(\as^2)$. At higher orders in $\as$, large
logarithms of $s/m_Q^2$ arise, potentially spoiling the convergence of
the perturbative series at high energies, $s \gg m_Q^2$.  It is
interesting to note that these are exactly the same logarithms which
occur in the resummation of jet multiplicities, where they appear as
logarithms of the jet resolution scale, $y_{\mathrm{cut}}$.  Here we
are able to study their effect without the arbitrariness of jet
algorithms.

In Ref.~[\ref{Mike}] the leading and next-to-leading logarithms were
summed to all orders in $\as$ yielding a result that is uniformly
reliable for all $s$. This gave the fraction of $Z^0$ decays that
contain a secondary charm or bottom quark pair, of
\be
\label{eq:mikeres} 
f_{\C}=1.349\%, \quad f_{\B}=0.177\%. 
\ee

\subsection{Experimental Progress}

Several experimental measurements of the secondary heavy quark
production rate have now been made. In Ref.~[\ref{OPAL1}], it was
extracted for charm quarks from a measurement of the $D^*$
fragmentation function, and found to be more than a factor of two
above the expectation of Ref.~[\ref{Mike}], although with large
systematic errors coming from uncertainty in the fragmentation
function of primary charm quarks. Refs.~[\ref{OPAL2}--\ref{ALEPH}]
made less model dependent measurements by selecting hard three-jet
events, which enhances the fraction of heavy quarks produced by the
gluon splitting mechanism. In general the measurements have been above
the predictions of Eq.~(\ref{eq:mikeres}), although within the range
allowed by variations in $\as$ and the quark mass.

The combined LEP values\footnote{For the bottom quark there has been
  only one measurement[\ref{DELPHI}], while for the charm quark, we
  have averaged the results of Refs.~[\ref{OPAL1},\ref{OPAL2}] and
  [\ref{ALEPH}], assuming that the systematic errors are
  uncorrelated.} obtained are,
\be
  f_\C = (2.44 \pm 0.43)\%, \quad
  f_\B = (0.22 \pm 0.13)\%.
\ee

Of course, in order to extract these results one must use some
theoretical input in order to separate the primary and secondary heavy
quark contributions. In the above experiments this theoretical input
was the shape of the secondary heavy quark multiplicity distribution
with respect to some event shape variable. For example,
Ref.~[\ref{ALEPH}] uses the shape of the multiplicity distribution
with respect to the jet mass difference. This is fitted to data,
extracting a value for the overall normalisation and therefore the
total rate. Since a full perturbative QCD calculation of this shape
was not available, Monte Carlo event
generators[\ref{JETSET}--\ref{ARIADNE}] were used for this theoretical
input.

\subsection{A New Calculation}

Clearly, it is desirable to calculate the resummed multiplicity,
retaining the dependence on the jet kinematics, and thereby allowing
the calculation of any event shape variable to be performed
numerically. This perturbative QCD calculation can be made accurate up
to next-to-leading logarithms whereas the Monte Carlo event generators
are only accurate to leading logarithms.  Furthermore, the leading
order, ${\cal O}(\as^2)$, can be included exactly, and the exact
kinematics can be used when calculating the required event shape.

Here I present the multiplicity of secondary heavy quarks as a
function of the {\em heavy jet mass}. I employ the `thrust-like'
definition of heavy jet mass, where the thrust axis is used to divide
each event into two hemispheres and the mass of the heavier hemisphere
is defined as the heavy jet mass. Clearly, the calculation can be
easily modified to any thrust-like event variable.

\section{The Secondary Heavy Quark Multiplicity as a Function of Heavy 
Jet Mass}

The leading order differential cross section for the production of
secondary heavy quarks, $\mathrm{\gamma^* \to q \bar q \qqb}$, is
easily calculated, and will not be discussed further here.
For the calculation of the logarithmic contribution, there are two
requirements to which we must conform. Firstly we must retain the
exact kinematics of the $q \bar qg$ production in order to be able to
accurately obtain the heavy jet mass for each event. We need not worry
about the exact kinematics of the heavy quarks, since the large
logarithms arise from the parts of phase space where they become
collinear and only the kinematics of the gluon need be considered
(although the exact kinematics of the heavy quarks are, of course,
included in the fixed order contribution). Secondly, we must include
all soft gluon emission from the light quarks and virtual gluon. These
emissions contribute to the heavy jet mass (by taking the light quarks
off mass-shell) and also provide large logarithms which must be summed
to all orders using the coherent branching formalism.

Bearing these considerations in mind, the differential multiplicity is 
taken to be,
\ba
n_{e^+e^-}^{\qqb}(M_H^2,Q^2;Q_0^2) = \int dx_1 \, dx_2 \, dk_1^2 \, 
dk_2^2 \, dk_g^2 \, \frac{x_1^2+x_2^2}{k_{\perp}^2/Q^2} 
f_q(k_1^2,k_{1 \, max}^2)f_q(k_2^2,k_{2 \, max}^2) \nn \\
n_g^{\qqb}(k_g^2,k_{\perp}^2;Q_0^2) \Theta(k_{\perp}^2-Q_0^2)
\delta(M_H^2-h(x_1,x_2,k_1^2,k_2^2,k_g^2)).
\label{eq:goveq} 
\ea
As usual, $x_1$ and
$x_2$ are the energy fractions of the light quark and antiquark
respectively, and $k_1$, $k_2$ and $k_g$ are the four-momenta of the
quarks and gluon. The maximum value of $k_i^2$, $i=1,2$, as
constrained by the phase space limits, is $k_{i \, max}^2$. Also,
$k_{\perp}^2$ is the transverse momentum (squared) of the virtual
gluon, given by,
\be
k_{\perp}^2 = (1-x_1+\epsilon_1-\epsilon_2)(1-x_2+\epsilon_2-\epsilon_1)Q^2,
\ee
where $Q^2$ is the centre-of-mass energy squared and $\epsilon_i$, ($i=1,2$)
are the rescaled (primary) quark masses (squared) which result from
the soft gluon emission. By retaining the exact kinematics of the quark,
antiquark and gluon, we are able to calculate the heavy jet mass
$h(x_1,x_2,k_1^2,k_2^2,k_g^2)$ exactly, thereby satisfying the first of our 
requirements.

The second requirement, that we should include all soft gluon
emission, has been achieved by the inclusion of the functions $f_q$
and $n_g^{\qqb}$. The function $f_q$ is the quark jet mass
distribution which has been calculated to next-to-leading logarithmic
accuracy in Ref.~[\ref{CTTW}]. More explicitly, $f_q(k^2,Q^2) \, dk^2$
is the probability that a quark, created at a scale $Q^2$ gives rise
to a jet with mass squared between $k^2$ and $k^2+dk^2$. This function
includes all soft gluon emission from the light quarks and sums, to all
orders, leading and next-to-leading logarithms of $k_{i \, max}^2/k_i^2$, 
$i=1,2$.

The function $n_g^{\qqb}$ is the gluon jet mass distribution weighted
by the heavy quark pair multiplicity. In other words,
$n_g^{\qqb}(k_g^2,k_{\perp}^2;Q_0^2) \, dk_g^2$ is the number of heavy
quark pairs within a gluon jet which was formed at a scale
$k_{\perp}^2$ and has a mass between $k_g^2$ and $k_g^2+dk_g^2$.  To
the accuracy required here this quantity must include leading and
next-to-leading logarithms of both $k_{\perp}^2/k_g^2$ and
$k_{\perp}^2/Q_0^2$.

One important simplification of the calculation is allowed by the
introduction of a heavy quark {\em effective} mass[\ref{Mike}],
\be
m_Q^*=\smfrac12 m_Q e^{5/6}.
\ee
Since the bottom quark mass cannot be neglected
we should include all mass effects in the $g \to b \bar b$ splitting.
This would lead to integrals of the from,
\be
\lim_{x^2 \to 0} \int_{x^2}^a \frac{dz}{z} \sqrt{1-\frac{x^2}{z}} \left( 1+
  \frac{x^2}{2z} \right) \log^{n-1}z = -\frac{1}{n} \log^n x^{* 2} +
{\cal O} \left (\log^{n-2} \right),
\ee
where $x$ is the rescaled heavy quark mass, $m_Q/ \surd s$, and $x^*$ is
the rescaled effective mass. However, exactly the same result is
obtained (to next-to-leading logarithmic accuracy) if the {\em
  massless} splitting function and the effective mass are used.
\be
\lim_{x^{*2} \to 0} \int_{x^{*2}}^a \frac{dz}{z} 
\log^{n-1}z = -\frac{1}{n} \log^n x^{* 2}.
\ee
Therefore by using this effective mass one can neglect the heavy quark
mass in the decay of the gluon while maintaining the correct leading
and next-to-leading logarithms. In Eq.~(\ref{eq:goveq}) this is
manifest as the resolution scale at which the heavy quarks are
resolved, $Q_0=2 \, m_Q^*$.

\subsection{The Multiplicity Weighted Mass Distribution}

It is more convenient to calculate the integrated distribution,
\be
N_g^{\qqb}(k^2,Q^2;Q_0^2)= \int_0^{k^2} dq^2 n_g^{\qqb}(q^2,Q^2;Q_0^2).
\ee
Physically this is the number of $\qqb$ pairs resolved in gluon jets
of mass squared {\em less than} $k^2$. It can be derived from
$N_g^g(k^2,Q^2;Q_0^2)$, the multiplicity of gluons within gluon jets
of mass squared less than $k^2$, by integrating over the kernel for
the splitting $\mathrm{g \to \qqb}$. Since we have introduced the
effective mass $m_Q^*$, the appropriate kernel is the massless
splitting kernel $P_{qg}$. $N_g^g$ has been derived in
Ref.~[\ref{us}]. The integration yields,
\be
N_g^{\qqb}(k^2,Q^2;Q_0^2)=F_g(k^2,Q^2) \left\{ \n_g^{\qqb}(k^2;Q_0^2) 
+ C_A \left(I^{\qqb}(k^2;Q_0^2)-I^{\qqb}(k^4/Q^2;Q_0^2) \right) \right\},
\label{eq:ngqq}
\ee
with,
\ba
I^{\qqb}(k^2;Q_0^2) &=& \Theta(z_k-z_0) \frac{4}{3b \, C_A} 
\left\{ \tN_1(z_0,z_k) + \left( 2(B-1)\frac{1}{z_0^2}- C\frac{1}{z_0z_k} 
\right) \n^+(z_0,z_k) \right. \nn \\
&& \left. - \frac{1}{z_k^2} \left( 2B-2-C \right) - \smfrac14 (C+2) \log 
\left( \frac{z_k^2}{z_0^2} \right) - \frac{C}{4} 
\left( 1- \frac{z_k^2}{z_0^2} \right) \right\}.
\ea
In the above, $\n_g^{\qqb}(k^2;Q_0^2)$ is the multiplicity of heavy
quarks pairs found in a gluon jet, regardless of the jet mass, and
$F_g(k^2,Q^2)$ is the probability that a gluon jet produced at a scale
$Q$ will have a mass squared below $k^2$. The notation used above is
standard and can be found in Ref.~[\ref{CDFW}]. $\n_1$ is defined as
$\n$ but with $B$ replaced by $B-1$.

Notice that Eq.~(\ref{eq:ngqq}) conforms with na{\"\i}ve expectations.
One might expect that the number of heavy quark pairs found within a
gluon jet created at a scale $Q^2$ and of mass below $k^2$ would be
simply the probability of finding a gluon jet of mass below $k^2$
multiplied by the number of heavy quark pairs within. Indeed, this is
the first term of our expression for $N_g^{\qqb}$, and the na{\"\i}ve
expectation requires only the addition of next-to-leading logarithmic
corrections.

The expression for $n_g^{\qqb}$ can now be trivially obtained from the 
differentiation of $N_g^{\qqb}$ with respect to the jet mass. 
This result is then matched to the fixed order result to avoid double
 counting.

\subsection{Calculation of the Background}

The background to secondary heavy quark production in $\mathrm{e^+ e^-}$
annihilation, i.e.\ primary heavy quark production, is estimated by
standard three-jet production, since the mass effects will be small. 
Care has been taken to include both the fixed order contribution and all
large leading and next-to-leading logarithms. As for the secondary heavy 
quarks, the logarithmic contribution is matched to fixed order to avoid 
double counting. The full result can be seen plotted in Fig.~\ref{fig:heavy}.

\section{Numerical Results}

For all the distributions we show, we concentrate on their shape,
normalised to the number of secondary heavy quarks, rather than on the
total rate. We use $\as=0.118$ and values of $1.2$ and $5$ GeV
for the charm and bottom quark masses respectively.

We present the heavy jet mass distribution for $\surd s=m_\Z$ in
Fig.~\ref{fig:heavy}.  This is closely related to the jet mass
difference, $M_H-M_L$, which was the event shape used to fit $f_c$ in
Ref.~[\ref{ALEPH}]. We see that the heavy jet mass provides a good
discriminator of events with secondary heavy quarks from the three-jet
background.

\begin{figure}[thbp]
\begin{minipage}[b]{.47\linewidth}
\centering\epsfig{file=sigandback.ps,angle=90,width=\linewidth,height=5cm}
\caption[heavy]{The multiplicity of primary and secondary heavy quark pairs 
  as a function of the heavy jet mass, normalised to the number of
  heavy quarks.}
\label{fig:heavy}
\end{minipage}
\hspace{0.78cm}
\begin{minipage}[b]{.47\linewidth}
\centering\epsfig{file=bqrate.ps,angle=90,width=\linewidth,height=5cm}
\caption[bquark]{The multiplicity of secondary bottom quark pairs as a 
function of the heavy jet mass, normalised to the number of $\Z^0$ 
decays.}
\label{fig:bquark}
\end{minipage}
\end{figure}

Of course, these shapes are dependent on the values chosen for the
parameters, $\Lambda_{\mathrm{QCD}}$ and $m_{\mathrm{Q}}$.  The effect
of varying the quark mass by $5\%$ (dotted) and
$\Lambda_{\mathrm{QCD}}$ by a factor of two (i.e.\ $\as$ by $10\%$)
(dash-dotted) is compared to the result with $\as=0.118$ and
$m_{\B}=5$ GeV (solid). Also shown is the contribution from the fixed
order term alone (dashed), demonstrating the importance of resumming
large logarithms.

\section{Event Generators}

The predictions from Monte Carlo event generators for $\surd s =
m_{\Z}$ can be seen in Fig.~\ref{fig:MC91}.  Since these models are
only formally accurate to leading logarithms and do not include the
exact matrix elements for $\mathrm{q\bar{q}Q\bar{Q}}$ production, our
calculation is more accurate and can be used to check them.  We see
that HERWIG and JETSET give similar predictions for the distribution
as well as for the rate and that ARIADNE peaks at a somewhat lower
heavy jet mass.  Following a suggestion of Ref.~[\ref{Mike}], later
versions of ARIADNE contain an option to veto gluon splitting with
$m_g>k_{\perp g}$. Adding this modification, ARIADNE's distribution is
more like the other models', but still somewhat different,
particularly at low jet masses.  Our results lie between the
unmodified ARIADNE and the other models.
\begin{figure}[thbp]
\begin{minipage}[b]{.47\linewidth}
\centering\epsfig{file=MC_91.ps,angle=90,width=\linewidth,height=5cm}
\caption[MC91]{The multiplicity of bottom quark pairs as a function of
  the heavy jet mass at $\surd s=m_{\Z}$.}
\label{fig:MC91}
\end{minipage}
\hspace{0.78cm}
\begin{minipage}[b]{.47\linewidth}
\centering\epsfig{file=MC_500.ps,angle=90,width=\linewidth,height=5cm}
\caption[MC500]{The multiplicity of bottom quark pairs as a function of
  the heavy jet mass at $\surd s=500$~GeV.}
\label{fig:MC500}
\end{minipage}
\end{figure}

Increasing the centre-of-mass energy, the relative importance of the
fixed order term is reduced and one gets a cleaner probe of the parton
evolution.  In Fig.~\ref{fig:MC500} we show the comparison at
$\surd s = 500$ GeV.  The modified version of ARIADNE is in even
better agreement with the other two models, while the unmodified
version is in good agreement with our calculation.  We therefore see
no evidence to support the claim of Ref.~[\ref{Mike}] that there is a
problem with ARIADNE.

\vspace{-1.5mm}
\section{Summary}
\vspace{-1mm}
A calculation of the multiplicity of heavy quarks from gluon splitting
in $\mathrm{e^+e^-}$ annihilation, as a function of the heavy jet
mass, has been presented. The shape of the result is similar to that
predicted by Monte Carlo event generators at the $\Z^0$, lying between
the different models, but in better agreement with ARIADNE at higher
energy.

\vspace{-1.5mm}
\section*{Acknowledgements}
\vspace{-1mm}
The author would like to thank the European Union for the award of a
Training and Mobility of Researchers grant to attend this conference.
The work presented in this talk was done in collaboration with 
Michael~H.~Seymour.

\vspace{-1.5mm}
\section*{References}
\vspace{-1mm}
\begin{enumerate}
\item\label{Mike}
  M.H. Seymour, \np{436}{163}{95}
\vspace{-2.5mm}
\item\label{Kniehl}
  B.A. Kniehl and J.H. K\"uhn, \np{329}{547}{90}
\vspace{-2.5mm}
\item\label{OPAL1}
  OPAL Collaboration, R. Akers {\em et al.}, \zp{67}{27}{95}
\vspace{-2.5mm}
\item\label{OPAL2}
  OPAL Collaboration, R. Akers {\em et al.}, \pl{353}{595}{95}
\vspace{-2.5mm}
\item\label{DELPHI}
  DELPHI Collaboration, P. Abreu {\em et al.}, \pl{405}{202}{97}
\vspace{-2.5mm}
\item\label{ALEPH}
  G. Hansper for the ALEPH Collaboration, Contribution to the XXVIII
  International Conference on High Energy Physics ICHEP Warsaw, Poland,
  24--31 July 1996, \mbox{Pa~05-065}
\vspace{-2.5mm}
\item\label{JETSET}
  T. Sj\"ostrand, \cpc{39}{347}{84}; \\ M. Bengtsson and T. Sj\"ostrand,
  \cpc{43}{367}{87}
\vspace{-2.5mm}
\item\label{HERWIG}
  G. Marchesini, B.R. Webber, G. Abbiendi, I.G. Knowles, M.H. Seymour
  and L. Stanco, \cpc{67}{465}{92}
\vspace{-2.5mm}
\item\label{ARIADNE}
  L. L\"onnblad, \cpc{71}{15}{92}
\vspace{-2.5mm}
\item\label{CTTW}
  S. Catani, L. Trentadue, G. Turnock and B.R. Webber,
  \np{407}{3}{93}
\vspace{-2.5mm}
\item\label{us}
  D.J. Miller and M.H. Seymour, `The Jet Multiplicity as a Function of
  Thrust', Rutherford Lab preprint RAL-TR-98-041, hep-ph/9805413
\vspace{-2.5mm}
\item\label{CDFW}
  S. Catani, Yu.L. Dokshitzer, F. Fiorani and B.R. Webber,
  \np{377}{445}{92}
\end{enumerate}

\end{document}